# Six-degrees-of-freedom test mass readout via optical phase-locking heterodyne interferometry

Xin Xu, Jinsong Liu, Henglin Mu, Yan Li, and Yidong Tan*

*Abstract*—Accurate position and posture measurements of the freely-falling test mass are crucial for the success of spaceborne gravitational wave detection missions. This paper presents a novel laboratory-developed test mass motion readout that utilizes quadrant photodetectors to measure the translation and tilt of a test mass. Departing from conventional methods like Zeeman effect or AOM frequency shift modulation, the readout system employs the phase locking of two lasers to generate the dual-frequency heterodyne source. Notably, the out-of-loop sensitivity of the phase locking reaches below 30 pm/Hz$^{1/2}$ within the frequency band of 1 mHz and 10 Hz. The system comprises three measurement interferometers and one reference interferometer, featuring a symmetric design that enables measurements of up to six degrees of freedom based on polarization-multiplexing and differential wavefront sensing. Ground-simulated experimental results demonstrate that the proposed system has achieved a measurement sensitivity of 4 pm/Hz$^{1/2}$ and 2 nrad/Hz$^{1/2}$ at 1 Hz, a resolution of 5 nm and 0.1 urad, a range of 200 um and 600 urad, respectively. These findings showcase the system's potential as an alternative method for precisely monitoring the motion of test masses in spaceborne gravitational wave detection missions and other applications requiring accurate positioning and multi-degrees-of-freedom sensing.

## I. Introduction

RECENTLY, several scientific missions aim to employ heterodyne interferometers in space, such as LISA, Taiji and Tianjin, opening a new window to the millihertz gravitational-wave universe [1-4]. In these missions, the critical focus is to construct an optical readout system for the relative distance measurement between two freely-falling test masses, also known as the gravitational reference sensor (GRS). To reach the goal sensitivity, the six degrees of freedom of the test mass should be monitored and pre-controlled down within the noise requirement at the interested frequency band [5]. LISA Pathfinder, launched by European Space Agency (ESA) in 2015, is seen as an important milestone in spaceborne gravitational wave detection development [6]. It combines the capacitive sensors and the optical heterodyne interferometry to measure the six-degrees-of-freedom motions of the test mass. The newest test results show that a measurement noise of 0.032 pm/Hz$^{1/2}$ of the sensitive-axis translation has been successfully achieved [7]. Though the capacitive sensors fail to provide the competitive picometer-level sensitivity for other five freedoms as the optical heterodyne interferometry, the optical-capacitive combination for the six-degrees-of-freedom measurement has been successfully verified and applied in the LISA Pathfinder.

Recently, inspired by the great success of LISA Pathfinder, many research groups continue to develop high-sensitivity optical readout for the multi-degrees-of-freedom measurement of the test mass. A five-degrees-of-freedom test mass readout via optical levels is designed and tested [8], and can achieve a measurement sensitivity of 300 nm/Hz$^{1/2}$ and 600 nrad/Hz$^{1/2}$ at frequencies between 10 mHz and 1 Hz. This alternative approach has the advantage of compactness, while its main weaknesses is the nanoscale sensitivity compared with the heterodyne interferometric readouts. In 2022, the research group from Tianqin proposed a heterodyne interferometric system based on differential wavefront sensing, which can measure six degrees of freedom of the test mass [9]. This all-optical sensing system theoretically provides a higher precision of the translation and tilt measurement than the optical-capacitive combined sensing method in LISA Pathfinder. Moreover, the capability to monitor six-degrees-of-freedom at the same time can obtain more useful information on the motion of the freely falling test mass.

In this paper, we construct a polarizing-multiplex heterodyne interferometric system for six-degrees-of-freedom test mass motion readout. Different from conventional heterodyne interferometers, phase locking loops are employed to generate the dual-frequency laser, which avoids the disadvantages of nonlinear errors using the Zeeman-effect dual-frequency laser [10-11], or laser power loss using a pair of acousto-optic modulators [12]. The optical bench has four heterodyne interferometers, and three of them aim to measure the translation and tilt of the test mass, while the other one is to provide a reference so as to greatly compensate the common-mode noises. The paper is organized into four main sections. First, the optical design and working principle are

Manuscript received Month xx, 2xxx; revised Month xx, xxxx; accepted Month x, xxxx. This work was supported by the National Key Research and Development Program of China (No. 2020YFC2200204 & No.2020YFC2200101). (*Corresponding author: Yidong Tan*).

The authors are with the Department of Precision Instruments, Tsinghua University, Beijing 100084, China (email: xx19@mails.tsinghua.edu.cn; tanyd@tsinghua.edu.cn).

introduced in Section II. Then, the experimental interferometric system is constructed and introduced in Section III, with which the performance results of out-of-loop sensitivity, the stability/resolution/range of six-degrees-of-freedom readout are shown in this Section. The conclusion and the future arrangements are presented in Section IV.

## II. DESIGN AND WORKING PRINCIPLE

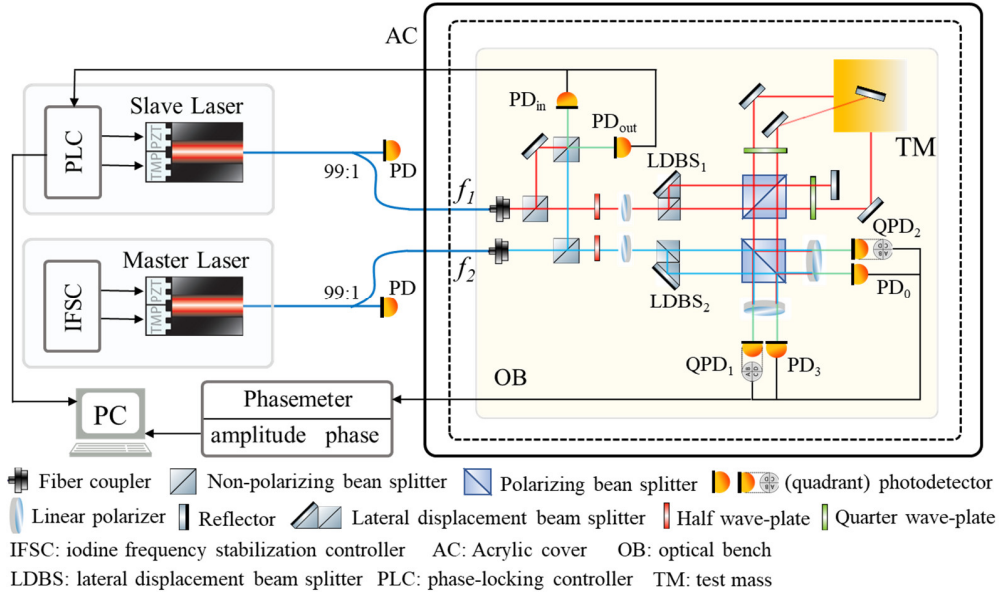

Fig. 1. Schematic diagram of the optical readout system.

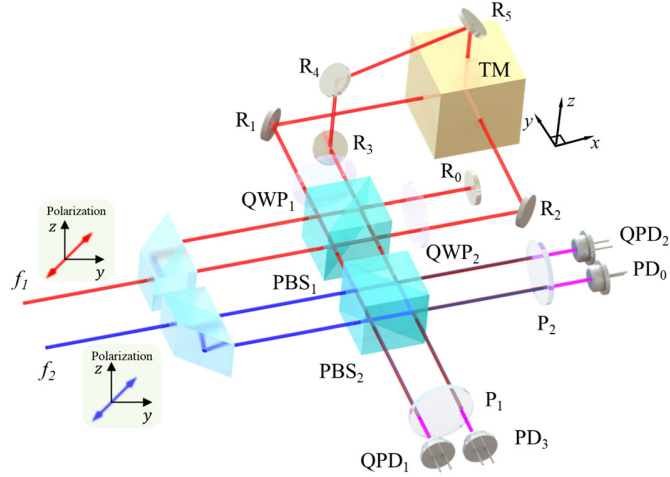

Fig. 2. The Specific path of the six-degrees-of-freedom readout. PBS: polarizing beam splitter; QWP: quarter wave-plate; P: polarizer; R: reflector; QPD: quadrant photodetector; PD: photodetector; TM: test mass;

Table.1 Detailed paths of the measurement beam and the reference beam for the two interferometers.

| Path1_m | $f_1 \to PBS_1 \to QWP_1 \to R_1 \to TM \to R_1 \to QWP_1 \to PBS_1 \to PBS_2 \to P_1 \to QPD_1$ | $QPD_1$ $\Delta x\ \theta x\ \theta y$ |
|---|---|---|
| Path1_r | $f_2 \to PBS_2 \to P_1 \to QPD_1$ | |
| Path2_m | $f_1 \to PBS_1 \to QWP_2 \to R_2 \to TM \to R_2 \to QWP_2 \to PBS_1 \to PBS_2 \to P_2 \to QPD_2$ | $QPD_2$ $\Delta y\ \theta x\ \theta z$ |
| Path2_r | $f_2 \to PBS_2 \to P_2 \to QPD_2$ | |
| Path3_m | $f_1 \to PBS_1 \to QWP_1 \to R_3 \to R_3 \to R_4 \to R_5 \to TM \to R_5 \to R_4 \to R_3 \to QWP_1 \to PBS_1 \to PBS_2 \to P_1 \to PD_3$ | $PD_3$ $\Delta z$ |
| Path3_r | $f_2 \to PBS_2 \to P_1 \to PD_3$ | |
| Path4_m | $f_1 \to PBS_1 \to QWP_2 \to R_0 \to QWP_2 \to PBS_1 \to PBS_2 \to P_2 \to PD_0$ | $PD_0$ reference |
| Path4_r | $f_2 \to PBS_2 \to P_2 \to PD_0$ | |

Fig. 1 illustrates the systemic configuration of six-degrees-of-freedom test mass readout via optical phase-locking heterodyne interferometry, including the heterodyne laser sources, the optical bench, and the data acquisition and processing part. Compared with traditional heterodyne interferometers, the proposed system utilizes a phase locking loop to generate the dual-frequency laser ($f_1$ and $f_2$) by controlling the two lasers locked to each other, which is also the same scheme in the future space-based gravitational wave detection missions [13]. The laser frequency tuning is driven by the PZT and thermal modulation from the fast and slow control output of the phase locking loop. The beat frequency difference can be changed by setting up the LO frequency. The detailed principle can be referred to this literature [14], and the performance can be effectively evaluated from the out-of-loop sensitivity, which is defined as the amplitude spectral density of the out-of-loop signal's phase [15]. The results of the difference between the out-of-loop phase and the in-loop phase are also given in the paper, noting that it does not have much practical significance but an indirect indication of the limiting performance using the phase locking in the current system.

Two fiber-injected beams ($f_1$ and $f_2$) with 45° linear polarization through a half waveplate and a polarizer, are then incident and split by the polarizing beam splitter (PBS$_{1,2}$). There are four interferometers in the proposed system, including three measurement interferometers and one reference interferometer. The detailed paths of the measurement beam and the reference beam for the four interferometers are listed in Table 1, and the 3D optical design for six-degrees-of-freedom readout is shown in Fig. 2. Taking one measurement interferometer as an example, the reference beam of horizontal linear polarization is incident onto the quadrant photodetector (QPD$_2$) after a polarizing beam splitter (PBS$_2$) and a polarizer (P$_2$). The measurement beam passes through the polarizing beam splitter (PBS$_1$), and the transmissive part is reflected by the reflector and the test mass. Twice passing through the quadrant wave-plate changes the polarizing direction of the backward beam, thus this beam can be reflected by the polarizing beam splitter (PBS$_1$). Then, the measurement beam is combined by the polarizing beam splitter (PBS$_2$) with the reference beam. The polarizer (P$_2$) keeps the same polarizing part and the beat signals are detected by the quadrant photodetector (QPD$_2$) with the active area divided into the quadrants A-B-C-D.

Based on differential wavefront sensing (DWS) and longitudinal pathlength sensing, a quadrant photodetector can measure three degrees of freedom [9,16]. In the proposed system, two of the measurement interferometers are used to measure five degrees of freedom, as the tilt along the z-axis is common for the two QPDs (QPD$_1$ and QPD$_2$), while another measurement interferometer is for the z direction displacement through the photodetector PD$_3$. The fixed reflector of the reference interferometer is used to provide a reference signal (PD$_0$) to subtract most common-mode pathlength fluctuations. Therefore, six-degrees-of-freedom motions can be measured, and the equations are expressed as:

$$\Delta x = \frac{\lambda}{4\pi}(\frac{\phi_{1A} + \phi_{1B} + \phi_{1C} + \phi_{1D}}{4} - \phi_0) \quad (1)$$

$$\Delta y = \frac{\lambda}{4\pi}(\frac{\phi_{2A} + \phi_{2B} + \phi_{2C} + \phi_{2D}}{4} - \phi_0) \quad (2)$$

$$\Delta z = \frac{\lambda}{4\pi}(\phi_3 - \phi_0) \quad (3)$$

$$\theta x \approx \frac{\lambda}{4\sqrt{2\pi}D} \cdot (\phi_{1A} + \phi_{1B} - \phi_{1C} - \phi_{1D}) \quad (4)$$

$$\theta y \approx \frac{\lambda}{4\sqrt{2\pi}D} \cdot (\phi_{2A} + \phi_{2B} - \phi_{2C} - \phi_{2D}) \quad (5)$$

$$\theta z \approx \frac{\lambda}{4\sqrt{2\pi}D} \cdot (\phi_{1A} - \phi_{1B} - \phi_{1C} + \phi_{1D})$$
$$\approx \frac{\lambda}{4\sqrt{2\pi}D} \cdot (\phi_{2A} - \phi_{2B} - \phi_{2C} + \phi_{2D}) \quad (6)$$

where $\phi_{ij}$ ($i = 0,1,2,3$ $j = A, B, C, D$) is the phase change calculated from the interference wavefront detected by the two QPDs (QPD$_1$ and QPD$_2$) or single-element photodetectors (PD$_0$ and PD$_3$). $\Delta x$, $\Delta y$ and $\Delta z$ represent the translations of the test mass, $\theta x$, $\theta y$ and $\theta z$ represent the three tilt degrees of freedom respectively. $\lambda$ is the measured laser wavelength and $D$ is the beam diameter on the quadrant photodetector. $\lambda/4\sqrt{2\pi}D$ is the first-order coefficient factor of the tilt measurement, the typical value of which is approximately ~$10^{-4}$ rad/rad [16].

III. PERFORMANCE TEST AND RESULTS

A. System

The experimental system contains three main parts, including two laser source unit, the optical bench of six-degrees-of-freedom interferometers and the phase readout subsystem. In this paper, we use a non-planar ring-oscillator (NPRO) type Nd:YAG laser (Coherent Mephisto 500NE) and an iodine frequency-stabilized laser (Coherent Prometheus P20NE) as the heterodyne laser sources. Part of them after the isolator is coupled into a single-mode polarization-maintaining fiber and injected into the optical bench by two fiber collimator (Thorlabs, CFP5-1064A). The output beams are split and the reflective parts are combined at one beam splitter to form a Mach-Zehnder interferometer. The in-loop and out-of-loop signals are sent into the phase locking controller (Liquid Instruments, Moku: Pro [17]). Two controlling signals of the fast and slow loops are generated to drive the PZT and thermal modulation of the free-running laser, thus the two lasers can be locked to each other with a fixed frequency difference. The frequency difference of the two lasers is set in a range of 5-25 MHz in our system, which is also the heterodyne frequency range used in the future space-based gravitational wave detection missions. After locking to each, the transmissive parts of the laser will pass a half wave-plate and a polarizer, and we can obtain two beams of $f_1$ and $f_2$ with 45° linear polarization for the use of six-degrees-of-freedom sensing.

The power of the input beams in the six-degrees-of-freedom sensing bench is approximately 5 mW and its waist diameter is about 1 mm. The optical bench is fixed on a marble table with an independent ground foundation and covered by an acrylic box to decrease the air perturbation. The beat signals are acquired by the photodetectors and sent to the phase demodulation subsystem. The subsystem consists of two phasemeters (Liquid Instruments, Moku: Pro), one of which is used for locking the two lasers, and the other is for the phase demodulation of multi-degrees-of-freedom measurements. It should be noted that six-degrees-of-freedom measurement needs a ten-channel phasemeter, which is currently not available and under development in our laboratory. Therefore, we can only measure three degrees of freedom at one time using the commercial phasemeter. In the experiments, all-optical components are carefully positioned symmetrically to minimize the unequal-arm length, therefore reducing the common mode noise of laser frequency fluctuations and ambient disturbance noise. In the experiments, the target is a quartz cube with six gold-plated surfaces, which is placed on a six-axis nano-positioning stage (Physik Instruments, P-562.6CD, resolution 1 nm and 100 nrad, range 200 μm and 1000 μrad).

## B. Performance results

### 1. Out-of-loop sensitivity using heterodyne phase locking

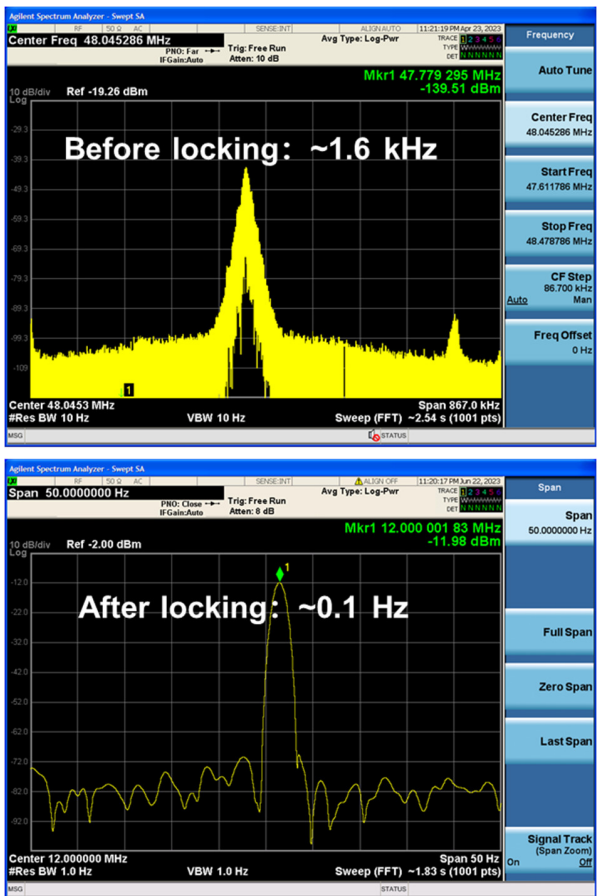

Fig. 3. Spectrum of the beat frequency signal before and after applying the phase locking.

We first test the basic performance of the beat frequency before and after applying the phase locking between two lasers, including the beat frequency linewidth, the frequency stability of time domain and amplitude spectral density. The results are shown in the Fig. 3 and Fig. 4. Before locking to the iodine frequency-stabilized laser, the beat frequency linewidth of two lasers is about 1.6 kHz, which is consistent with the laser manual instructions [18]. When it is locked, the linewidth of the beat frequency is greatly suppressed to be narrowed than 1 Hz, as shown in the Fig. 3 recorded by a spectrum analyzer (Agilent, N9020A). The stability improvement of the beat frequency is also evident from the Fig. 4 results. As one of the locked lasers is iodine frequency-stabilized, the frequency fluctuations are mainly from the free-running laser (Coherent Mephisto 500NE), which is about 10 MHz/Hz$^{1/2}$ at 1 mHz, while the beat signal will get an excellent stability of less than 1 Hz drift in several hours and 0.1 Hz/Hz$^{1/2}$ at 1 mHz after applying the phase locking.

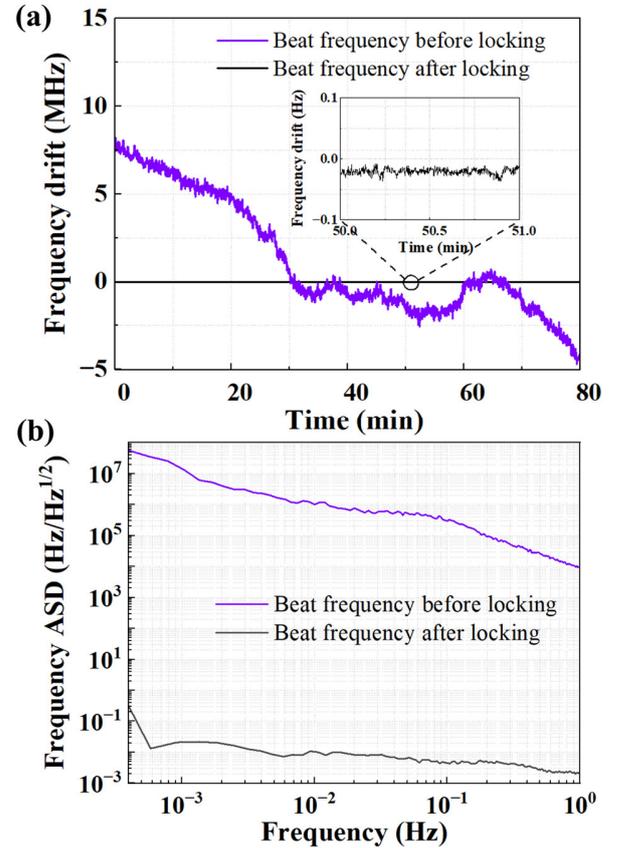

Fig. 4. Stability of the beat frequency before and after applying the phase locking. (a) time domain (b) amplitude spectral density.

Furthermore, a few experiments are contacted to characterize the phase-locking performance. We compared the out-of-loop sensitivity by using two different output signals from the phase-locking controller as the error signal, one is the direct phase output and the other one is the frequency offset (the difference between the actual beat frequency and the preset heterodyne frequency) [17]. The results are shown in Fig. 5, indicating that both methods can reach below 30 pm/Hz$^{1/2}$ above 20 mHz. Nevertheless, there is a bit difference between two methods as for the out-of-loop sensitivity, which needs more experiments to analyze the reasons. Another experiment is to study the effects of the scale factor (the conversion of the phase to analog voltage, the unit is mV/rad) and the locking heterodyne frequency for the phase locking performance, and the results are shown in Fig. 6 and Fig. 7, respectively. Obviously, larger the scale factor is, better performance the out-of-loop sensitivity can reach in the phase locking experiments. Nevertheless, when the scale factor gets larger than 160 mV/rad, the locking state can not last stably for more than 4 hours. Therefore, in the next experiment of changing the heterodyne frequency, the scale factor is set as 160 mV/rad. From the Fig. 7 results, it tells that the locking heterodyne frequency has nearly no effect on the out-of-loop sensitivity, and the phase locking performance can reach below 30 pm/Hz$^{1/2}$ at the frequency band of 1 mHz and 10 Hz, with a heterodyne frequency of 5-25 MHz.

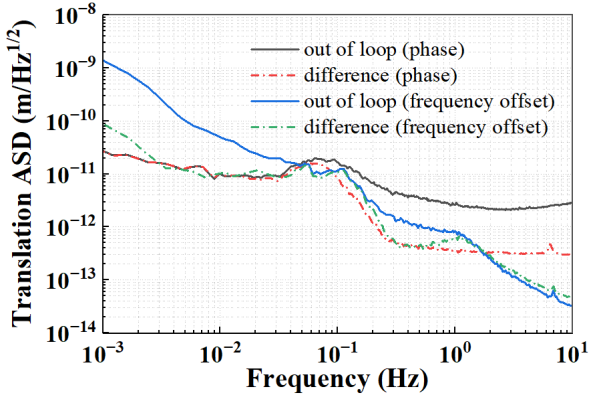

Fig. 5. The effective displacement sensitivity using different output as the error signal. 'phase' means the error signal sent to the phase locking controller is the phase output of the in-loop signal. 'frequency offset' means the error signal is the difference value between the measured frequency by the phasemeter and the preset frequency.

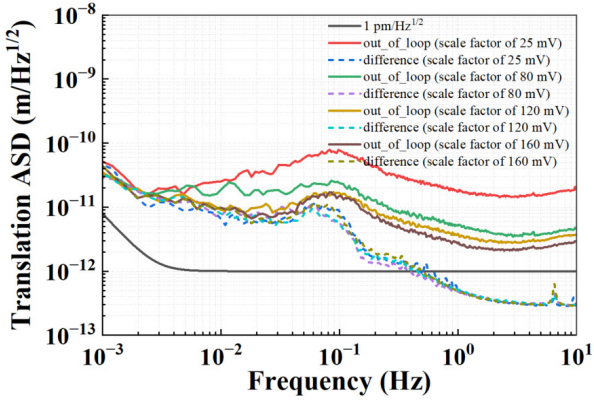

Fig. 6. The effective displacement sensitivity using different output as the error signal.

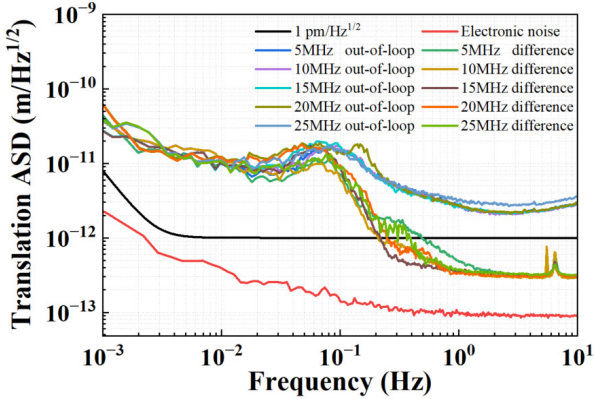

Fig. 7. The effective displacement sensitivity with different preset heterodyne frequency.

*2. Stability of six-degrees-of-freedom measurement*

Then we evaluate the six-degrees-of-freedom measurement stability of the optical interferometric system. The amplitude spectral density (ASD) curves of the translation and tilt are illustrated in Fig. 8. The results show that the measurement noise can reach 4 pm/Hz$^{1/2}$ and 2 nrad/Hz$^{1/2}$ at 1 Hz. Learning from many previous researches on the measurement noise of heterodyne interferometers, we make a preliminary identification on the possible noise causes in the different frequency bands. For the current optical system, the noise below 1 mHz is mainly caused by the temperature variation, and it can be corrected by a precise temperature control during the test or a detrend data processing of the test results. The noise above 1 Hz is from the vibration of the test bench, which can be sharply reduced by a differential detection of common-path measurement. At the frequency band of 1 mHz – 1 Hz the sensitivity limit is currently owing to the optical path difference fluctuations of the air perturbations. Nevertheless, the above analysis is preliminary and a further study is needed to identify the specific sources of noise and conduct a quantitative analysis of the interferometer's performance.

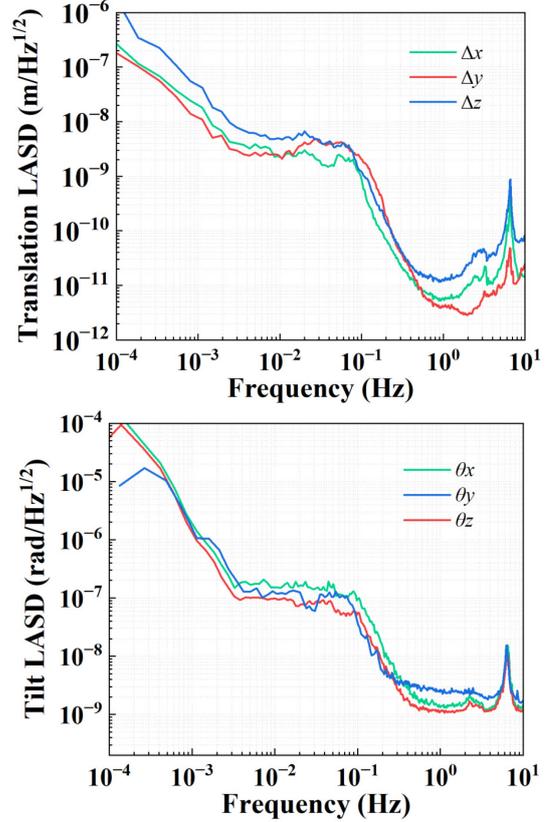

Fig. 8 Measurement results of six-degrees-of-freedom motions. Up: ASD of translation $\Delta x$, $\Delta y$ and $\Delta z$; Down: ASD of tilt $\theta x$, $\theta y$ and $\theta z$.

*3. Resolution & range*

To test the minimum incremental motion of the six-degrees-of-freedom measurement, we use a commercial nano-positioning stage as an actuation stage. The experimental resolution of the translation is 5 nm and of the tilt is 100 nrad, as shown in Fig. 9.

For the translation resolution experiments, the effective resolution of the used actuation stage is 1 nm. However, the constructed interferometer is unable to distinguish such tiny displacement. To figure out this point, a short-term drift of the target with different installation methods is tested. The stability of the target fixed on the stage is found to be about 3 nm, while the stability of the target fixed directly on the bench is about 0.4 nm, as shown in Fig. 10. Therefore, it is reasonable to assume that the constructed interferometer has a resolution better than 5 nm, as the current stage lacks the stability required for such tiny displacement resolution experiments. If we use the standard deviation of a short-term drift (10s) as the resolution, the minimum incremental motion of the translation is approximately 45 pm.

For the tilt resolution experiments, the effective resolution of the used actuation stage is 100 nrad, and the constructed interferometer can clearly distinguish such tilt. Like the analysis of the translation, we also test the short-term tilt drift with different installation conditions. Owing to the differential wavefront sensing, the vibration can be effectively eliminated, as the sensing beams pass through almost the same path. The minimum incremental motion of the tilt is about 10 nrad calculated from the standard deviation of a 10s tilt drift.

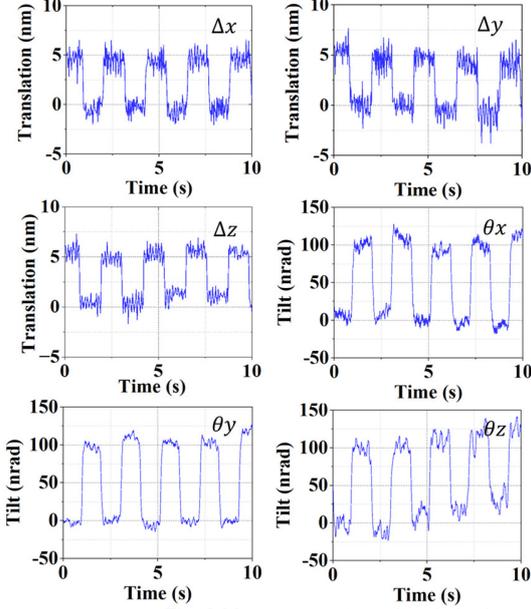
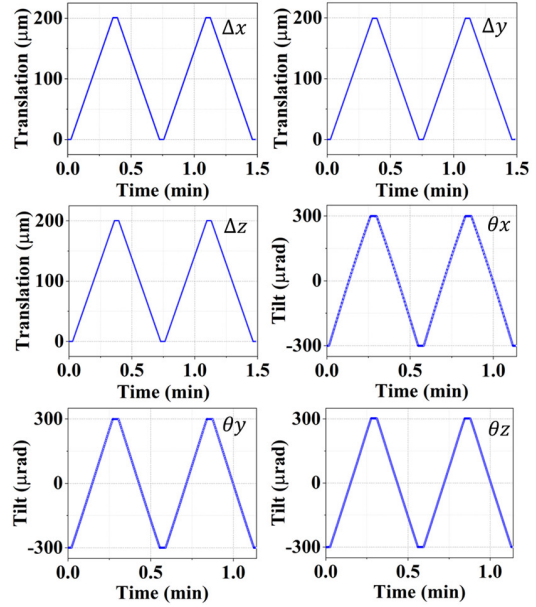

Fig. 9 Measurement results of six-degrees-of-freedom motions. Left: resolution; Right: range.

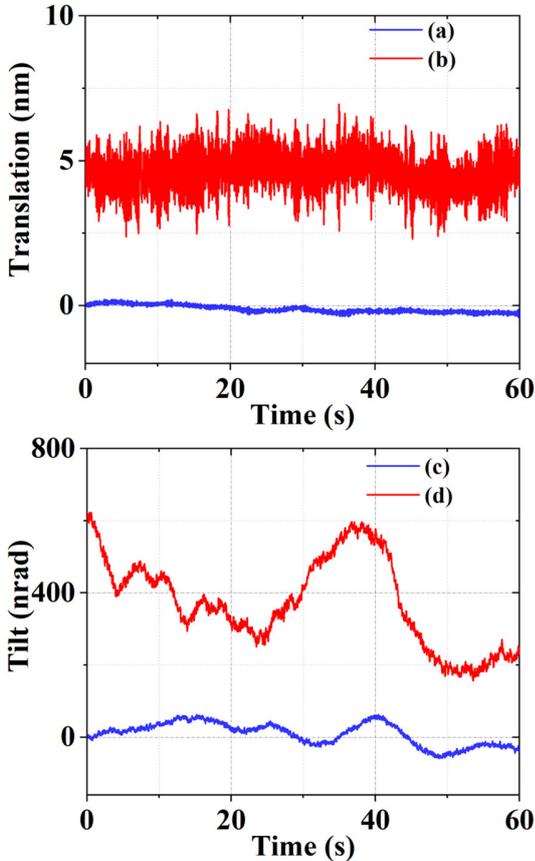

Fig. 10 Stability results of short-term drift. (a)Translation measurement when the target is fixed directly on the bench (b)Translation measurement when the target is fixed on the nano-positioning stage (c)Tilt measurement when the target is fixed directly on the bench. (d)Tilt measurement when the target is fixed on the nano-positioning stage.

In addition to the resolution experiments, the range of six-degrees-of-freedom interferometers is also tested. A back-and-forth full-range motion of translation and tilt is carried out using a nano-positioning stage, and the tested results are shown in Fig. 9. Due to the range limitations of the motion stage, the measurement ranges for the translation and the tilt are 200 μm and 1000 μrad, respectively. Nevertheless, these ranges are sufficient for spaceborne gravitational wave detection applications.

## IV. CONCLUSION

This paper proposes and presents the construction of an optical heterodyne interferometric system for the measurement of translation and tilt. The optical design can be extended for up to six-degrees-of-freedom measurement based on polarization multiplexing and differential wavefront sensing. The common-path and symmetric design enable a low measurement noise level. The stability, resolution, sensing range of the translation and tilt are tested. The translation measurement noise is 3 pm/ $Hz^{1/2}$ at 1 Hz and the tilt measurement noise is 2 nrad/$Hz^{1/2}$ at 1 Hz. The laboratory development of such optical heterodyne interferometers for the multiple-dimension motions measurement can serve as a high-precision reference sensor and for controlling the freely-falling test mass in spaceborne gravitational wave detection.

In the next stage, we plan to construct an optical-bonding bench and place the developed testbed in a vacuum chamber. To suppress the measurement noise, the vibration should be passively isolated and the operating temperature needs to keep stable with multi-layer thermal shields. We believe that controlling the test environment can effectively improve the sensitivity of the translation and tilt measurement for the test mass [19-21]. Although there is still a lot of work to be done, we hope that this laboratory development of heterodyne interferometric system will provide a potential solution or relative technology verifications to the translation and tilt measurement of the freely falling test mass in the future spaceborne gravitational wave detection and other applications that need multi-degree-of-freedom measurement and precise positioning.